\documentclass[12pt]{article}
\usepackage{geometry}
\usepackage{setspace}
\usepackage{exscale}
\usepackage{relsize}
\usepackage{extarrows} 
\usepackage{subfigure}
\usepackage[numbers,sort&compress]{natbib}
\usepackage{times}
\usepackage{graphicx}
\usepackage{subfigure}
\usepackage{graphics}
\newcounter{lastnote}

\makeatletter 
\@addtoreset{equation}{section}
\makeatother  

\begin{document}
\title{Holographic Schwinger effect with a rotating probe D3-brane}

\author{Hao Xu$^{1,2}$ \thanks{xuhao1107@gmail.com}, M. Ilyas$^3$ \thanks{ilyas\_mia@yahoo.com} and Yong-Chang Huang$^{2,4}$ \thanks{ychuang@bjut.edu.cn}\\
$^1$ Deep Space Exploration Laboratory, Beijing 100043, China\\
$^2$ Institute of Theoretical Physics,\\
Beijing University of Technology, Beijing 100124, China\\
$^3$ Institute of Physics, Gomal University,\\
Dera Ismail Khan, 29220, KP, Pakistan\\
$^4$ CCAST(WorldLab.), P.O. Box 8730, 100080, Beijing, China}

\date{}
\maketitle

\begin{center}
\section*{Abstract}
\end{center}
~~~~This paper, among other things, talks about possible research on the holographic Schwinger effect with a rotating probe D3-brane. We discover that for the zero temperature case in the Schwinger effect, the faster the angular velocity and the farther the distance of the test particle pair at D3-brane, the potential barrier of total potential energy also grows higher and wider. This paper shows that at a finite temperature, when $S^5$ without rotation is close to the horizon, the Schwinger effect fails because the particles remain in an annihilate state, which is an absolute vacuum state. However, the angular velocity in $S^5$ will avoid the existence of an absolute vacuum near the horizon. For both zero and finite temperature states, the achieved results completely agree with the results of the Dirac-Born-Infeld (DBI) action. So the theories in this paper are consistent. All of these show that these theories will play important roles in future pair production research.

\clearpage
\newpage
\section{Introduction}~~~~The Schwinger effect \citep{schwinger1951gauge} is a famous non-perturbational phenomenon in quantum electrodynamics. A strong electric field converts the virtual electron-positron pair into real particles. Many strong external fields will produce a neutral particle pair from high-dimensional spacetime, such as strings and D-branes \citep{gorsky2002schwinger}. As a result, the Schwinger effect has an objective existence and is required for a thorough understanding of the vacuum structure and non-perturbative aspects of string theory and quantum field theories \citep{sato2013potential}.

To become real particles, virtual particles must be able to overcome some potential from external field energy. As a quantum process, the potential energy is closely related to the quantum tunnelling effect \citep{sato2013holographic}. Assuming that virtual particles are separated by a distance, which is simply analysed via the Coulomb potential, then the expression of potential contains two possible quantum processes. The critical electric field $E_c$ is a key factor in these two processes. When $E < E_c$, the potential energy barrier exists and quantum tunnelling can occur, so the vacuum stays stable. Whereas if $E > E_c$, the real particle production rate will no longer have the exponentially suppressed property, and the vacuum will become catastrophically unstable.

How to get the critical electric field is an important problem. The external electric field's critical value is assumed to be in QED (Gorsky 2002 schwinger).Unfortunately, it might not be observed from the pair-production rate computed in QED \citep{affleck1982pair}.

So, it's interesting to think about the Schwinger effect in the context of holography, and \citep{gorsky2002schwinger,semenoff2011holographic,zhang2016holographic} has started a similar study.
If the system is the $\mathcal{N}=4$ super Yang-Mills theory, it can calculate the critical field accurately, and the result agrees with the DBI result \citep{semenoff2011holographic}. The extension of the magnetic field has also been completed \citep{bolognesi2013comments,sato2013holographic2}. However, if we estimate the critical field by using Coulomb potential to calculate the AdS/CFT instruction \citep{das2010probe}, there would be some deviation from both the string world-sheet result and the DBI result \citep{rey2001macroscopic,maldacena1998wilson}. Yoshiki Sato and Kentaroh Yoshida investigated a static potential by evaluating the classical action of a probe D3-brane in AdS spacetime, and the resulting critical field agrees perfectly with the DBI result \citep{sato2013potential,zhang2011dbi}.

Separating one D3-brane from a parallel stack of N coincident D3-branes yields the Higgsing of $U(N+1)$ to $U(1)\times U(N)$ \citep{semenoff2011holographic}. The $AdS_5\times S^5$ background replaces the D3-brane stake \citep{sato2013holographic2} in the large N limit. The probe D3-brane is fixed at $r=r_0$ in AdS space. R. G. Cai \emph{et al}. demonstrate that the super Yang-Mills corresponding to rotating D3-branes are in the Higgs phase, and that we can set the probe to have angular momentum in some $\phi$-direction \citep{rey2001macroscopic,cai1999dynamics,gubser1999thermodynamics,brandhuber1999wilson,freedman2000continuous,minahan1998quark}. One approach is to connect rotating open strings to the probe bares \citep{sonnenschein2014rotating}. 

In this paper, we would like to investigate the holographic Schwinger effect by setting the probe D3-brane to have an angular velocity and calculating its potential function. In Section II, we analyse the potential at zero temperature and then compare it to the DBI result. In Section III, we investigate the effect at low temperatures and discover a link between separate distance and D3-brane position. The critical field $E_c$ also completely agrees with the DBI result. Section IV is the discussion, summary, and conclusion.

\section{Research on potential at zero temperature}
\subsection{Setup}
~~~~The Schwinger effect is a phenomenon that occurs in quantum field theory, where a strong electric field can create pairs of particles out of the vacuum. In the case of the D3-brane, this effect can occur when a test particle pair is separated by a distance and has a high angular velocity.
At zero temperature, the energy of the system is very low, and the behavior of the particles is dominated by their quantum mechanical properties. In this limit, the potential barrier that must be overcome to create a particle-antiparticle pair becomes very high and wide as the distance between the particles and their angular velocity increase.
The brane world model is a theoretical framework that arises from string theory, which is a high-energy physics theory. However, the behavior of particles in the low-energy limit can provide insight into the behavior of gravity and other fundamental forces on large scales, which is relevant to the brane world model.
By studying the behavior of particles in the low-energy limit, we can understand the properties of the brane world model, such as the behavior of gravity on a large scale, the existence of extra dimensions, and the possible existence of other fundamental particles beyond those in the Standard Model of particle physics.
Therefore, even though the brane world model originates from high-energy physics, by studying the behavior of particles in the low-energy limit, we can still gain insight into the behavior of the model and its implications for our understanding of the universe.

We should set up the needed background spacetime. First, the metric in the Poincare coordinates of $\mathrm{AdS_5  \times S^5}$ is
\begin{eqnarray} \label{2-1}
	\mathrm{d}s^2
	& = &
	\frac
		{r^2}
		{L^2}
	\eta_{\mu\nu}
	\mathrm{d}x^{\mu}
	\mathrm{d}x^{\nu}
	+\frac{L^2}{r^2} \mathrm{d}r^2
	+L^2 \mathrm{d} \Omega_5^2.
\end{eqnarray}
where $L$ denotes the $AdS$ radius and $\mathrm{d}\Omega_5^2$ denotes $S^5$, the coordinates $x^\mu(\mu=0,...,3)$ denotes a four-dimensional slice for each $r$. For simplicity, when we consider rotation, without losing generality, we choose one rotational motion direction, i.e. $\mathrm{d}\Omega_5=\mathrm{d}\phi$. The $\eta_{\mu\nu}$ is the metric, given as,
\begin{equation*}
  \eta_{\mu\nu}= diag(-1,1,1,1) \qquad\qquad (Lorentzian)
\end{equation*}
\begin{equation*}
  \eta_{\mu\nu}= diag(1,1,1,1) \qquad\qquad (Euclidean)
\end{equation*}

At finite temperature, the D3-brane black hole is characterized by a Hawking temperature, which is proportional to the surface gravity at the event horizon. When the temperature is non-zero, there is a thermal bath of particles that can interact with the D3-brane.
In the absence of rotation, the particles created by the Schwinger effect can annihilate with particles in the thermal bath, resulting in a net-zero effect. This is because there is no mechanism to break the symmetry between the created particles and the particles in the thermal bath.
However, when the D3-brane is rotating, it can create an effective potential that lifts the energy of the particles created by the Schwinger effect. This potential can prevent the created particles from annihilating with particles in the thermal bath, allowing the effect to persist even at finite temperature.
The physical mechanism behind this is related to the centrifugal force generated by the rotation of the D3-brane. This force creates an effective potential that depends on the angular momentum of the particles, which can be strong enough to prevent annihilation with the thermal bath.
In summary, the rotation of the D3-brane creates an effective potential that prevents the particles created by the Schwinger effect from annihilating with the thermal bath at finite temperature. This is due to the centrifugal force generated by the rotation, which creates a potential that depends on the angular momentum of the particles.

\subsection{Potential research}
~~~~In this work, the critical electric-field is argued with the DBI action in the Lorentzian signature, while the static potentials are computed in the conventional way with the Euclidean signature.

The next step is the computation of the classical solution from the string world-sheet. We choose the Euclidean signature and the Nambu-Goto action of string is
\begin{eqnarray} \label{2-2}
	S
	&=&
	T_F
	\int
		\mathrm{d}\tau
	\int
		\mathrm{d}\sigma
	\sqrt{\mathrm{det}G_{ab}}.
\end{eqnarray}
where the induced metric $G_{ab}(a,~b=0,1)$ is given by $G_{ab}=\frac{\partial x^\mu}{\partial \sigma^a}\frac{\partial x^\nu}{\partial \sigma^b}g_{\mu\nu}$.
String's world-sheet coordinates of string are $\sigma^a=(\tau,~\sigma)$. For our assumption, it is handy to work on the static gauge \citep{jorjadze2012bosonic}, which is given by
\begin{eqnarray} \label{2-4}
	x^0=\tau,
	&~~~~&
	x^1=\sigma.
\end{eqnarray}
For the classical solution, suppose that the radial direction only depends on $\sigma$ and the angular direction only depends on $\tau$,
\begin{eqnarray}\label{2-5}
	r=r(\sigma),
	&~~~~&
	\phi=\phi(\tau).
\end{eqnarray}

We can get Lagrangian density ($\mathcal{L}$)through induced metric $G_{ab}$ under the ansatz above, so
\begin{eqnarray} \label{2-7}
	G_{00}
	& = &
	\frac
		{\partial x^0}
		{\partial \sigma^0}
	\frac
		{\partial x^0}
		{\partial \sigma^0}~
	g_{00}
	+
	\frac
		{\partial x^\phi}
		{\partial \sigma^0}
	\frac
		{\partial x^\phi}
		{\partial \sigma^0}~
	g_{\phi\phi}
	\nonumber \\
& = & \frac{r^2}{L^2} + L^2\dot{\phi}^2,
~~~~~~\dot{\phi}\equiv\frac{\partial\phi}{\partial\tau},
\nonumber \\
G_{01} & = &
G_{10} ~ = ~ 0,
\nonumber \\
G_{11} & = & \frac{\partial x^1}{\partial \sigma^1}\frac{\partial x^1}{\partial \sigma^1}~g_{11}
+ \frac{\partial x^r}{\partial \sigma^1}\frac{\partial x^r}{\partial \sigma^1}~g_{rr}
\nonumber \\
& = & \frac{r^2}{L^2} + \frac{L^2}{r^2}(\frac{\partial r}{\partial\sigma})^2.
\end{eqnarray}
Then
\begin{eqnarray}\label{2-8}
	\mathcal{L}
	& = &
	\sqrt{
		(\frac
			{\mathrm{d}r}
			{\mathrm{d}\sigma}
		)^2
		+
		\frac{r^4}{L^4}
		+
		r^2\dot{\phi}^2
		+
		\frac{L^4}{r^2}
		(\frac
			{\mathrm{d}r}
			{\mathrm{d}\sigma}
		)^2
		\dot{\phi}^2
	}.
\end{eqnarray}
Because $\mathcal{L}$ (Lagrangian density) doesn't depend on $\sigma$ explicitly, it is easy to prove that
\begin{eqnarray}\label{2-9}
	\frac
		{\partial\mathcal{L}}
		{\partial(\partial_\sigma r)}
	\partial_\sigma r
	-
	\mathcal{L}
\end{eqnarray}
is conserved. Putting Eq.\eqref{2-8} into Eq.\eqref{2-9}, we achieve that
\begin{eqnarray}\label{2-10}
	\frac
		{
			\frac
				{\mathrm{d}r}
				{\mathrm{d}\sigma}
			+
			\frac
				{L^4}
				{r^2}
			\dot{\phi}^2
			\frac
				{\mathrm{d}r}
				{\mathrm{d}\sigma}
		}
		{
			\sqrt{
				(\frac
					{\mathrm{d}r}
					{\mathrm{d}\sigma})^2
				+
				\frac
					{r^4}
					{L^4}
				+
				r^2\dot{\phi}^2
				+
				\frac
					{L^4}
					{r^2}
				(\frac
					{\mathrm{d}r}
					{\mathrm{d}\sigma}
				)^2
				\dot{\phi}^2}
		}
	~
	\frac
		{\mathrm{d}r}
		{\mathrm{d}\sigma}
	~~~~~~~~~~~~
	\nonumber \\
	~~~~~~~~~
	-
	\sqrt{
		(\frac
			{\mathrm{d}r}
			{\mathrm{d}\sigma}
		)^2
		+
		\frac
			{r^4}
			{L^4}
		+
		r^2\dot{\phi}^2
		+
		\frac
			{L^4}
			{r^2}
		(\frac
			{\mathrm{d}r}
			{\mathrm{d}\sigma}
		)^2
		\dot{\phi}^2}
		& = &
		C,
\end{eqnarray}
we can obtain that
\begin{eqnarray}\label{2-11}
	-\frac
		{\frac{r^4}{L^4}+r^2\dot{\phi}^2}
		{\sqrt{(\frac{\mathrm{d}r}{\mathrm{d}\sigma})^2
		+\frac{r^4}{L^4}
		+r^2\dot{\phi}^2
		+\frac{L^4}{r^2}(\frac{\mathrm{d}r}{\mathrm{d}
		\sigma})^2\dot{\phi}^2}}
	& = &
	C.
\end{eqnarray}
Considering the boundary condition, when $\sigma=0$,
\begin{eqnarray}\label{2-12}
\frac
	{\mathrm{d}r}
	{\mathrm{d}\sigma}
	=0
&,&
r=r_c
~~(r<r_c).
\end{eqnarray}
The boundary condition is associated with the particular D3-brane on the string world-sheet, ending on it. It is to put a probe D3-brane at an intermediate position $r=r_c$.

Substituting Eq.\eqref{2-12} into Eq.\eqref{2-11} due to the boundary condition, we can find the conserved constant
\begin{eqnarray}\label{2-14}
	C
	& = &
	-\sqrt{
		\frac
			{r_c^4}
			{L^4}
		+r_c^2\dot{\phi}^2
	}.
\end{eqnarray}
Putting Eq.\eqref{2-14} into Eq.\eqref{2-11}, we have a differential equation
\begin{eqnarray}\label{2-16}
	\frac
		{\mathrm{d}r}
		{\mathrm{d}\sigma}
	& = &
	\frac
		{r^2}
		{L^2r_c}
	\sqrt{
		\frac
			{(r^2-r_c^2)(r^2+r_c^2+L^4\dot{\phi}^2)~}
			{r_c^2+L^4\dot{\phi}^2}
	}.
\end{eqnarray}
\begin{figure}[ht]
\centering
\includegraphics[width=7cm]{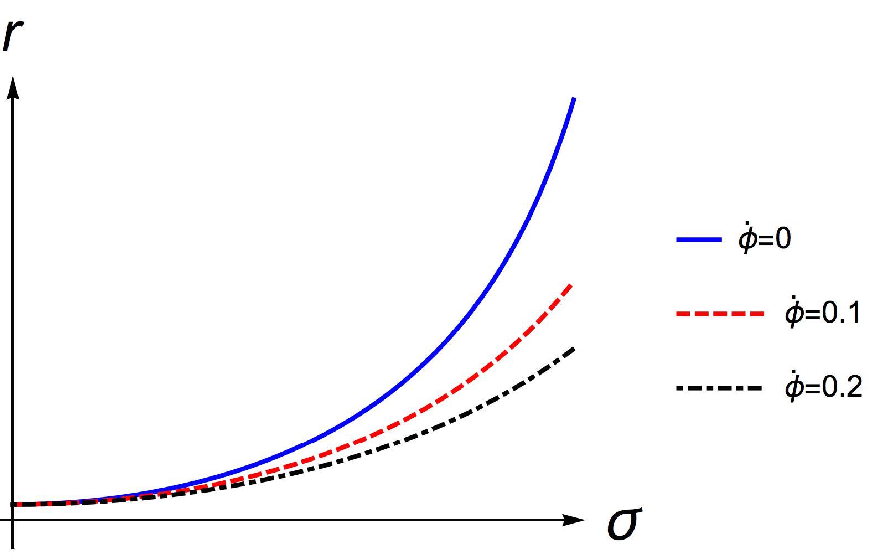}
\caption{The diagram shows the relationship between different angular velocities and different $r$ and $\sigma$. When we vary the angular velocity while keeping $r$ constant, we discover that the $\sigma$ increases as the angular velocity $\phi$ increases.}
\label{fig-1}
\end{figure}
We plot the numerical result of the differential equation \eqref{2-16} in Fig.\ref{fig-1}. When we change the angular velocity and select the same $r$, we find that $\sigma$ will increase with the increase of the angular velocity $\dot{\phi}$. In addition, the equation of motion \eqref{2-16} can also be achieved through taking Lagrangian density \eqref{2-11} into the Lagrange equation directly. We can obtain the same calculation results this way. Here, we choose the energy conservation method, is to simplify the calculation above. And it can decrease the order of the differential equation from 2-order to 1-order.
\begin{figure}[ht]
\centering
\includegraphics[width=8cm]{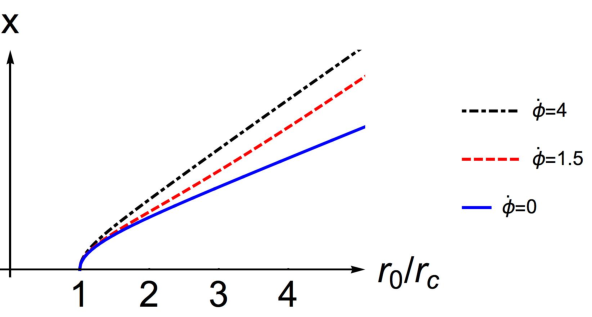}
\caption{When the angular velocities $\dot{\phi} = 0$, $\dot{\phi} = 1.5$ and $\dot {\phi} = 4$ respectively, the figure shows that the relation between the separate length $x$ of the test particle pair on the probe brane and $r_0/r_c$.}
\label{fig-2}
\end{figure}
We should calculate the separate length $x$ of test particles in the probe brane, we get the expression of $x$ by integrating the differential equation \eqref{2-16},
\begin{eqnarray}\label{2-17}
	\mathlarger{x}
	& = &
	 \mathlarger{2\int_{r_c}^{r_0}}
		\mathrm{d}r~
		\frac
			{L^2r_c}
			{r^2}
		\sqrt{
			\frac
				{r_c^2+L^4\dot{\phi}^2}
				{(r^2-r_c^2)(r^2+r_c^2+L^4\dot{\phi}^2)~}
	},
\end{eqnarray}
where $r_0$ denotes the intermediate position between $r_c$ and $\infty$. Then we can calculate and plot the relationship between the separate length $x$ and $r_0/r_c$, which is just shown in Fig.\ref{fig-2}.
\begin{figure}[ht]
\centering
\includegraphics[width=7cm]{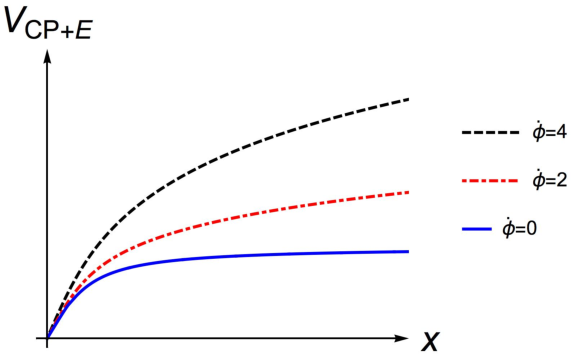}
\caption{When the angular velocity $\dot{\phi} = 0,2,4$, respectively, the figure shows that the relation between the separate length $x$ of test particle pair on the probe brane and the sum of Coulomb potential and Energy $V_{\mathrm{CP+E}}$. }
\label{fig-3}
\end{figure}

According to the previous setting, now we can achieve the modified form of the sum of Coulomb potential and Energy,
\begin{eqnarray}\label{2-18}
	V_{\mathrm{CP+E}}
	& = &
	2~T_F\int_0^{x/2}
		\mathrm{d}\sigma ~\mathcal{L}
	\nonumber \\
	\nonumber \\
	& = &
	2~T_F\mathlarger{\int_{r_c}^{r_0}}
		\mathrm{d}r~
		\frac
			{r^2+L^4\dot{\phi}^2}
			{
			  \sqrt{
			  	\left(
					r^2-r_c^2
				\right)
				\left(
					r^2+r_c^2+L^4\dot{\phi}^2
				\right)
			  }~
			},
\end{eqnarray}
where $T_F$ represents the fundamental string tension ($T_F=1/2\pi \acute{\alpha}$). The results of the different angular velocities corresponding to $V_{\mathrm{CP+E}}$ are showed in Fig.\ref{fig-3}.

\subsection{Calculating the critical field by DBI action}
~~~~In this part, we will investigate the critical field using DBI action. We also use the same spacetime background in Section 2.1 and use the Lorentz signature. Choosing the rotation direction as $\phi$, the metric form is given by
\begin{eqnarray}\label{2-19}
	\mathrm{d}s^2
	& = &
	\frac
		{r^2}
		{L^2}
	\mathrm{d}t^2
	+\frac
		{L^2}
		{r^2}
	\mathrm{d}r^2
	-\frac
		{r^2}
		{L^2}
	\sum_{i=1}^3
		\left(\mathrm{d}x^i\right)^2
	+L^2\mathrm{d}\phi^2
\end{eqnarray}

And then we consider the DBI action of a rotating probe D3-brane \citep{sonnenschein2014rotating} on the AdS background with a constant world-volume electric-field $E$ \citep{sato2013holographic2}. The rotating probe D3-brane is fixed at $r = r_0$. So the DBI action is written into the form
\begin{eqnarray}\label{2-20}
	S_{\mathrm{DBI}}
	& = &
	-T_{\mathrm{D3}}\int
			\mathrm{d}^4x
			\sqrt{-\mathrm{det}(G_{\mu\nu}+\mathcal{F}_{\mu\nu})~},
\end{eqnarray}
where $T_{\mathrm{D3}}$ is the D3-brane tension
\begin{eqnarray}\label{2-21}
	T_{\mathrm{D3}}
	& = &
	\frac
		{1}
		{g_\mathrm{s}(2\pi)^3\alpha^{\prime2}}.
\end{eqnarray}
We use Eq.\eqref{2-19} to compute the induced metric $G_{\mu\nu}$
\begin{eqnarray}\label{2-22}
	G_{00}
	 =
	\left(\frac
		{\partial x^0}
		{\partial x^0}
	\right)^2g_{00}
	+\left(\frac
		{\partial \phi}
		{\partial x^0}
	\right)^2g_{\phi\phi}
	& = &
	\frac
		{r^2}
		{L^2}
	+L^2\dot{\phi}^2,
	~~~~~~\dot{\phi} = \frac
					{\partial \phi}
					{\partial x^0},
	\nonumber \\
	G_{11}
	 =
	-\frac
		{r^2}
		{L^2},~~~
	G_{22}
	&=&
	-\frac
		{r^2}
		{L^2},~~~
	G_{33}
	=
	-\frac
		{r^2}
		{L^2}.
\end{eqnarray}
Now let us consider the $\mathcal{F}_{\mu\nu}$ term. There is a factor $2\pi\alpha^\prime$ relative to $\mathcal{F}_{\mu\nu} = 2\pi\alpha^\prime F_{\mu\nu}$ \citep{zwiebach2004first}, thus we have
\begin{eqnarray}\label{2-23}
	G_{\mu\nu}+\mathcal{F}_{\mu\nu}
	& = &
	\left(
		\begin{array}{cccc}
			r^2/L^2+L^2\dot{\phi}^2 & 2\pi\alpha^\prime E_1 & 2\pi\alpha^\prime E_2 & 2\pi\alpha^\prime E_3
			\\
			-2\pi\alpha^\prime E_1 & -r^2/L^2 & 0 & 0
			\\
			-2\pi\alpha^\prime E_2 & 0 & -r^2/L^2 & 0
			\\
			-2\pi\alpha^\prime E_3 & 0 & 0 & -r^2/L^2
		\end{array}
	\right).
\end{eqnarray}
It implies that
\begin{eqnarray}\label{2-24}
	\mathrm{det}(G_{\mu\nu}+\mathcal{F}_{\mu\nu})
	& = &
	-
	\left(
		\frac
			{r^2}
			{L^2}
	\right)^4
	\left(
		1
		-\frac
			{(2\pi\alpha^\prime)^2L^4}
			{r^4}
		E^2
		+
		\frac
			{L^4}
			{r^2}
		\dot{\phi}^2
	\right).
\end{eqnarray}
Taking Eq.\eqref{2-24} into the DBI action \eqref{2-20} and make the probe D3-brane located at $r=r_0$, we get
\begin{eqnarray}\label{2-25}
	S_{\mathrm{DBI}}
	& \xlongequal{r=r_0} &
	-T_{\mathrm{D3}}
	\frac
		{r_0^4}
		{L^4}
	\int
		\mathrm{d}^4x
		\sqrt{
			1
			-\frac
				{(2\pi\alpha^\prime)^2L^4}
				{r^4}
			E^2
			+\frac
				{L^4}
				{r^2}
			\dot{\phi}^2~
		}.
\end{eqnarray}
In order to avoid the action \eqref{2-25} being ill-defined, we need
\begin{eqnarray}\label{2-26}
	1
	-\frac
	{(2\pi\alpha^\prime)^2L^4}
	{r^4}
	E^2
	+\frac
		{L^4}
		{r^2}
	\dot{\phi}^2
	& \geq &
	0.
\end{eqnarray}
So the range of electric field is
\begin{eqnarray}\label{2-27}
	E
	& \leq &
	\frac
		{r_0^2}
		{L^2}
	T_F
	\sqrt{
		1
		+\frac
			{L^4}
			{r_0^2}
		\dot{\phi}^2
	}.
\end{eqnarray}
We can finally get the critical field value, which is given by
\begin{eqnarray}\label{2-28}
	E_c
	& = &
	\frac
		{r_0^2}
		{L^2}
	T_F
	\sqrt{
		1
		+\frac
			{L^4}
			{r_0^2}
		\dot{\phi}^2
	}.
\end{eqnarray}

\subsection{Total potential}
\begin{figure}[ht]
\centering
\includegraphics[width=7cm]{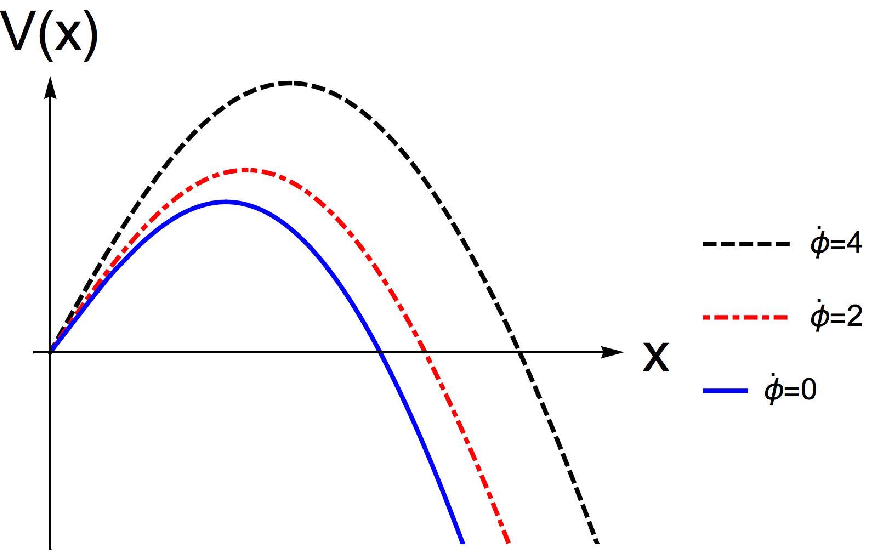}
\caption{When $\alpha=0.8$, if the angular velocity $\dot{\phi}=0$, $\dot{\phi}=2$?$\dot{\phi}=4$ respectively, the relation between total potential $V_{tot}$ and the separate distance $x$ of test particle pair on the rotating D3-brane is shown.}
\label{fig-4}
\end{figure}

~~~~We compute the total potential by turning on an electric field $E$ along the $x^1$ direction \citep{sato2013holographic2}. For simplicity, we introduce a variable dimensionless parameter
\begin{eqnarray}\label{2-29}
	\alpha
	&\equiv&
	\frac
		{E}
		{E_c}.
\end{eqnarray}
According to Eq.\eqref{2-17} and Eq.\eqref{2-18}, the total potential $V_{tot}$ is
\begin{eqnarray}\label{2-30}
	V_{tot}
	& = &
	V_{\mathrm{CP+E}}-Ex
	\nonumber \\
	& = & 2~T_F\mathlarger{\int_{r_c}^{r_0}}
		\mathrm{d}r~
		\frac
			{r^2+L^4\dot{\phi}^2}
			{
			  \sqrt{
			  	(r^2-r_c^2)(r^2+r_c^2+L^4\dot{\phi}^2)
			  }~
			}
	\nonumber \\
	\nonumber \\
	&~~&
	-~2\alpha~
	\frac
		{r_0^2}
		{L^2}
	T_F
	\sqrt{
		1
		+\frac
			{L^4}
			{r_0^2}
		\dot{\phi}^2
	}
	 \mathlarger{\int_{r_c}^{r_0}}
	 	\mathrm{d}r
		\frac
			{L^2r_c}
			{r^2}
		\sqrt{
			\frac
				{r_c^2+L^4\dot{\phi}^2}
				{(r^2-r_c^2)(r^2+r_c^2+L^4\dot{\phi}^2)~~~}
	}.~~~~
\end{eqnarray}
When $\alpha=0.8$, if the angular velocity $\dot{\phi}=0$, $\dot{\phi}=2$ and $\dot{\phi}=4$ respectively, the relation between total potential $V_{tot}$ and the separate distance $x$ of the test particle pair on the rotating D3-brane is shown in Fig.\ref{fig-4}.

As a quantum effect, the generation of real particle pairs in the Schwinger effect depends on quantum tunneling. In order to become a real particle pair, a virtual electron-positron pair need to obtain more energy than the rest energy of the particles from the external electric field \citep{sato2013holographic}. Eq.\eqref{2-30} is the difference between the potential and the energy from the external electric field. If we revisit the quantum mechanics, we will find that the existence of a potential barrier can make the production. The rate can be suppressed exponentially, causing the vacuum to become stable.

\begin{figure}[!ht]
	\centering
	\subfigure[]{\includegraphics[width=5cm]{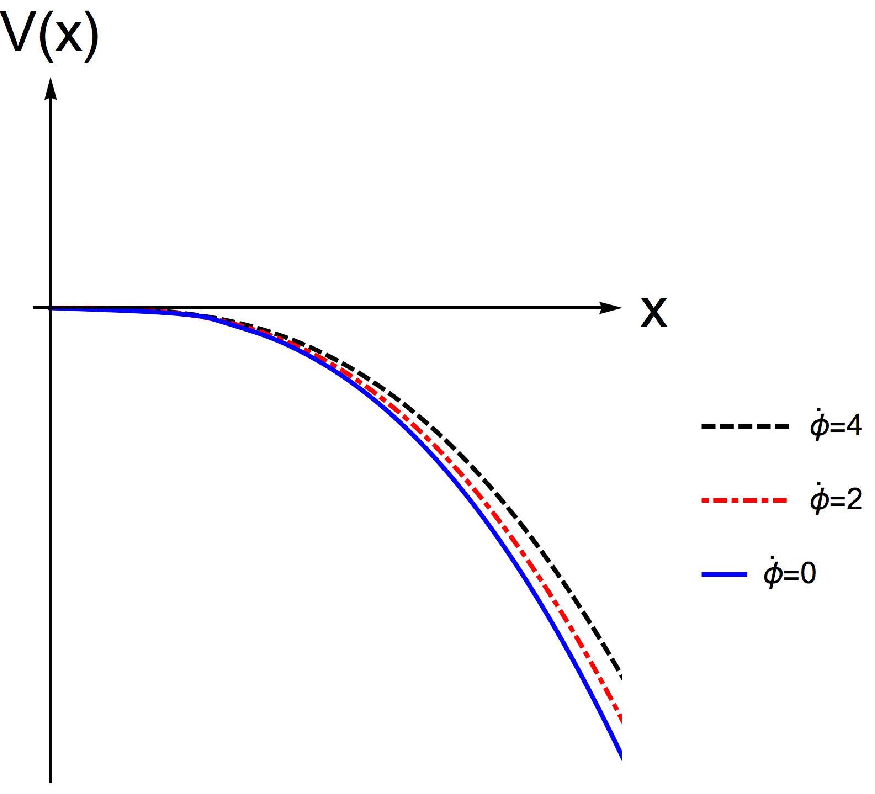}}~~~~ \subfigure[] {%
	\includegraphics[width=5cm]{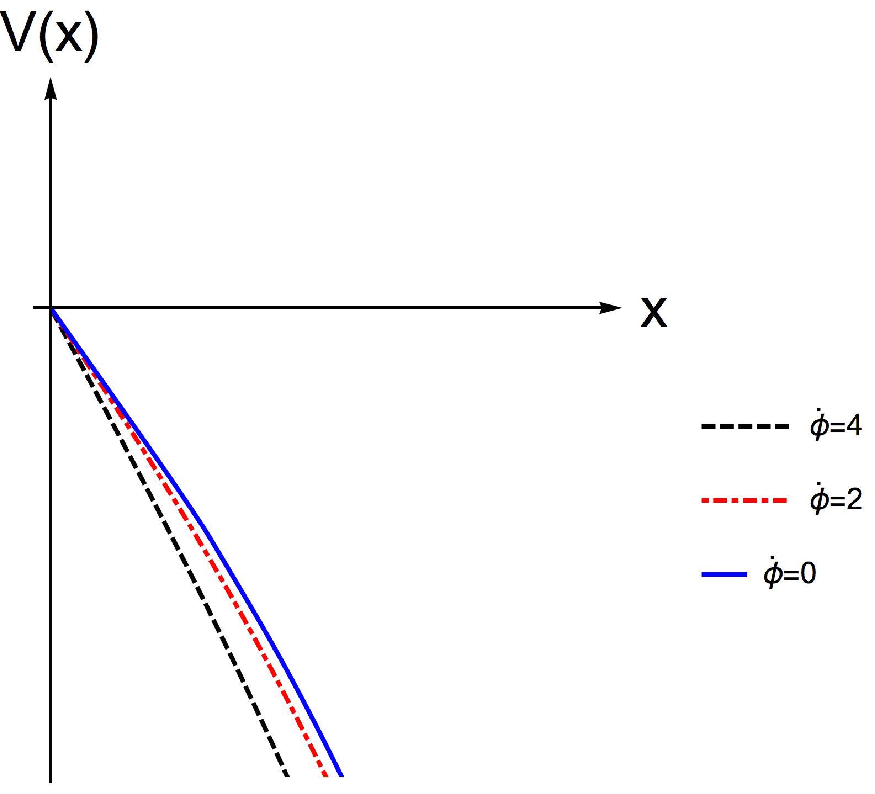}}
	\caption{(a) When $\alpha=1$ ($E=E_c$), the potential barrier vanishes, real particle production rate no longer is exponentially suppressed, the vacuum begins to become unstable. (b) When $\alpha=1.2$ ($E>E_c$), the potential barrier already vanishes. There is not any quantum tunneling effect any more.}
	\label{fig-5}
\end{figure}

Aiming to different values of $\alpha$ (corresponding to different electric field $E$), the pair production effects are also different. When $\alpha = 1$ ($E = E_c$) and $\alpha = 1.2$ ($E > E_c$) respectively, we plot the results in Fig.\ref{fig-5}. If $\alpha = 1$, the barrier just disappears and it can be strictly proved, i.e. we can calculate the derivative of $V_ {tot}$ and fix the location at $x = 0$, thus we have
\begin{eqnarray}
	\left.\frac
		{\mathrm{d}V_{tot}}
		{\mathrm{d}x}
	\right|_{x=0}
	& = &
	\left(1-\alpha\right)T_F
	\sqrt{
		1
		+\frac
			{L^4}
			{r_0^2}
		\dot{\phi}^2~
	}.
\end{eqnarray}
When $\alpha=1$, the potential barrier vanishes, the production rate of real particles no longer is exponentially suppressed, the vacuum begins to become unstable. This critical field $E=E_c$ completely agrees with the DBI result \eqref{2-28}.

\section{Research on potential at finite temperature}
\subsection{Setup}
~~~~In this section, we will consider the finite temperature case. First we should introduce an AdS planar black hole \citep{marolf2009black}. The metric with a Lorentz signature is
\begin{eqnarray}\label{3-1}
	\mathrm{d}s^2
	& = &
	\frac
		{r^2}
		{L^2}
	\left(1-\frac{r_h^4}{r^4}\right)
	\mathrm{d}t^2
	+\frac
		{L^2}
		{r^2}
	\left(1-\frac{r_h^4}{r^4}\right)^{-1}
	\mathrm{d}r^2
	-\frac
		{r^2}
		{L^2}
	\sum_{i=1}^3
	\left(\mathrm{d}x^i\right)^2
	+L^2 \mathrm{d} \Omega_5^2.
\end{eqnarray}
The horizon is fixed at $r=r_h$ and the temperature of black hole is \citep{sato2013potential}
\begin{eqnarray}\label{3-2}
	T
	& = &
	\frac
		{r_h}
		{\pi L^2}.
\end{eqnarray}
This temperature is dual to gauge theory \citep{freedman2000continuous}.

\subsection{Calculating the critical field by DBI action}
~~~~We need to consider the DBI action of a rotating probe D3-brane on the AdS black hole background with a constant world-volume electric field $E$. The rotating probe D3-brane is fixed at $r = r_0$ and we follow the assumption as
\begin{eqnarray}\label{3-3}
	r_h<r_c<r_0.
\end{eqnarray}
So the DBI action is written as
\begin{eqnarray}\label{3-4}
	S_{\mathrm{DBI}}
	& = &
	-T_{\mathrm{D3}}
		\int
			\mathrm{d}^4x
			\sqrt{
				-
				\mathrm{det}
				(
					G_{\mu\nu}
					+
					\mathcal{F}_{\mu\nu}
				)
			~},
\end{eqnarray}
where $T_{\mathrm{D3}}$ is the brane tension and the expression of $T_{\mathrm{D3}}$ is also \eqref{2-21}.

We use Eq.\eqref{3-1} to compute the induced metric $G_{\mu\nu}$,
\begin{eqnarray}\label{3-5}
	G_{00}
	=
	\left(\frac
		{\partial x^0}
		{\partial x^0}
	\right)^2g_{00}
	+\left(\frac
		{\partial \phi}
		{\partial x^0}
	\right)^2g_{\phi\phi}
	& = &
	\frac
		{r^2}
		{L^2}
	\left(
		1-\frac
			{r_h^4}
			{r^4}
	\right)
	+L^2\dot{\phi}^2,
	~~~~~~\dot{\phi} = \frac
					{\partial \phi}
					{\partial x^0},
	\nonumber \\
	G_{11}
	~=~
	G_{22}
	&=&
	G_{33}
	~=~
	-\frac
		{r^2}
		{L^2}.
\end{eqnarray}
Considering the $\mathcal{F}_{\mu\nu}$ term again, we have
\begin{eqnarray}\label{3-6}
	G_{\mu\nu}+\mathcal{F}_{\mu\nu}
	& = &
	\left(
		\begin{array}{cccc}
			r^2(1-r_h^4/r^4)/L^2+L^2\dot{\phi}^2 & 2\pi\alpha^\prime E_1 & 2\pi\alpha^\prime E_2 & 2\pi\alpha^\prime E_3
			\\
			-2\pi\alpha^\prime E_1 & -r^2/L^2 & 0 & 0
			\\
			-2\pi\alpha^\prime E_2 & 0 & -r^2/L^2 & 0
			\\
			-2\pi\alpha^\prime E_3 & 0 & 0 & -r^2/L^2
		\end{array}
	\right).
\end{eqnarray}
Therefore,
\begin{eqnarray}\label{3-7}
	\mathrm{det}(G_{\mu\nu}+\mathcal{F}_{\mu\nu})
	& = &
	-\left(\frac
		{r^2}
		{L^2}
	\right)^4
	\left(1
	-\frac
		{r_h^4}
		{r^4}
	+\frac
		{L^4}
		{r^2}
	\dot{\phi}^2
	-\frac
		{\left(2\pi\alpha^\prime\right)^2L^4}
		{r^4}
	E^2
	\right).
\end{eqnarray}
Taking Eq.\eqref{3-7} into the DBI action \eqref{3-4} and making the probe D3-brane located at $r=r_0$, we can achieve that
\begin{eqnarray}\label{3-8}
	S_{\mathrm{DBI}}
	& \xlongequal{r=r_0} &
	-T_{\mathrm{D3}}
	\frac
		{r_0^4}
		{L^4}
	\int
		\mathrm{d}^4x
		\sqrt{
			1
			-\frac
				{r_h^4}
				{r_0^4}
			+\frac
				{L^4}
				{r_0^2}
			\dot{\phi}^2
			-\frac
				{(2\pi\alpha^\prime)^2L^4}
				{r_0^4}
			E^2
		}.
\end{eqnarray}
In order to avoid the action \eqref{3-8} being ill-defined, we need
\begin{eqnarray}\label{3-9}
	1
	-\frac
		{r_h^4}
		{r_0^4}
	+\frac
		{L^4}
		{r_0^2}
	\dot{\phi}^2
	-\frac
		{(2\pi\alpha^\prime)^2L^4}
		{r_0^4}
	E^2
	& \geq &
	0.
\end{eqnarray}
So the range of electric field is
\begin{eqnarray}\label{3-10}
	E
	& \leq &
	\frac
		{r_0^2}
		{L^2}
	T_F
	\sqrt{
		1
		-\frac
			{r_h^4}
			{r_0^4}
		+\frac
			{L^4}
			{r_0^2}
		\dot{\phi}^2
	}.
\end{eqnarray}
We can finally get the critical field value at finite temperature, i.e.,
\begin{eqnarray}\label{3-11}
	E_c
	& = &
	\frac
		{r_0^2}
		{L^2}
	T_F
	\sqrt{
		1
		-\frac
			{r_h^4}
			{r_0^4}
		+\frac
			{L^4}
			{r_0^2}
		\dot{\phi}^2
	}.
\end{eqnarray}
It can be seen that the critical electric field $E_c$ depends on the temperature and the angular velocity.

\subsection{Potential analysis}
~~~~Just like in the zero temperature case, we need to analyze the classical solution of the string world-sheet. The signature is Euclidean. The assumptions do not change, and there are some modifications in $G_{ab}$ induced metric. The Nambu-Goto action is
\begin{eqnarray}\label{3-12}
	S
	&=&
	T_F
	\int
		\mathrm{d}\tau
	\int
		\mathrm{d}\sigma
	\sqrt{
		\mathrm{det}G_{ab}
	~},
\end{eqnarray}
where $G_{ab}(a,~b=0,1)$ is induced metric, i.e. $G_{ab}=\frac{\partial x^\mu}{\partial \sigma^a}\frac{\partial x^\nu}{\partial \sigma^b}g_{\mu\nu}$.
The world-sheet coordinates of string are also $\sigma^a=(\tau,~\sigma)$ and it is convenient to work on the static gauge, which is given by Eq.\eqref{2-4}. The radial direction only depends on $\sigma$ and angular direction only depends on $\tau$, that is, Eq.\eqref{2-5}.

We should calculate the induced metric $G_{ab}$ under the ansatz above. First,
\begin{eqnarray}\label{3-17}
	G_{00}
	& = &
	\frac
		{\partial x^0}
		{\partial \sigma^0}
	\frac
		{\partial x^0}
		{\partial \sigma^0}~
	g_{00}
	+
	\frac
		{\partial x^\phi}
		{\partial \sigma^0}
	\frac
		{\partial x^\phi}
		{\partial \sigma^0}~
	g_{\phi\phi}
	\nonumber \\
	& = &
	\frac
		{r^2}
		{L^2}
	\left(
		1-\frac
			{r_h^4}
			{r^4}
	\right)
	+
	L^2\dot{\phi}^2,
	~~~~~~
	\dot{\phi}
	\equiv
	\frac
		{\partial\phi}
		{\partial\tau},
	\nonumber \\
	G_{01} & = & G_{10} ~= ~0,
	\nonumber \\
	G_{11}
	& = &
	\frac
		{\partial x^1}
		{\partial \sigma^1}
	\frac
		{\partial x^1}
		{\partial \sigma^1}~
	g_{11}
	+
	\frac
		{\partial x^r}
		{\partial \sigma^1}
	\frac
		{\partial x^r}
		{\partial \sigma^1}~
	g_{rr}
	\nonumber \\
	& = &
	\frac
		{r^2}
		{L^2}
	+
	\frac
		{L^2}
		{r^2}
	\left(
		1-\frac
			{r_h^4}
			{r^4}
	\right)^{-1}
	\left(\frac
		{\partial r}
		{\partial\sigma}
	\right)^2.
\end{eqnarray}
Then we can obtain the Lagrangian density
\begin{eqnarray}\label{3-18}
	\mathcal{L}
	& = &
	\sqrt{
	\left(
		\frac
			{\mathrm{d}r}
			{\mathrm{d}\sigma}
	\right)^2
	+
	\frac
		{r^4}
		{L^4}
	\left(
		1-\frac
			{r_h^4}
			{r^4}
	\right)
	+
	r^2\dot{\phi}^2
	+
	\frac
		{L^4}
		{r^2}
	\left(
		1-\frac
			{r_h^4}
			{r^4}
	\right)^{-1}
	\left(
		\frac
			{\mathrm{d}r}
			{\mathrm{d}\sigma}
	\right)^2
	\dot{\phi}^2
	}.~~~~
\end{eqnarray}
Because $\mathcal{L}$ doesn’t depend on $\sigma$ explicitly, which implies that the conserved quantity is
\begin{eqnarray}\label{3-19}
	\frac
		{\partial\mathcal{L}}
		{\partial(\partial_\sigma r)}
	\partial_\sigma r
	-
	\mathcal{L}
	& = &
	C_1.
\end{eqnarray}
Putting Eq.\eqref{3-18} into Eq.\eqref{3-19}, we can get
\begin{eqnarray}\label{3-20}
	\frac
		{\frac
			{\mathrm{d}r}
			{\mathrm{d}\sigma}
		+
		\frac
			{L^4}
			{r^2}
		\left(
			1-\frac
				{r_h^4}
				{r^4}
		\right)^{-1}
		\dot{\phi}^2
		\frac
			{\mathrm{d}r}
			{\mathrm{d}\sigma}
		}
		{\sqrt{
			\left[
				1+
				\frac
					{L^4}
					{r^2}
				\left(
					1-\frac
						{r_h^4}
						{r^4}
				\right)^{-1}
				\dot{\phi}^2
			\right]
			\left(
				\frac
					{\mathrm{d}r}
					{\mathrm{d}\sigma}
			\right)^2
			+
			\frac
				{r^4}
				{L^4}
			\left(
				1-\frac
					{r_h^4}
					{r^4}
			\right)
			+
			r^2\dot{\phi}^2~
			}
		~}
	~
	\frac
		{\mathrm{d}r}
		{\mathrm{d}\sigma}
	~~~~~~~
	\nonumber \\
	-~
	\sqrt{
		\left[
			1+
			\frac
				{L^4}
				{r^2}
			\left(
				1-\frac
					{r_h^4}
					{r^4}
			\right)^{-1}
			\dot{\phi}^2
		\right]
		\left(
			\frac
				{\mathrm{d}r}
				{\mathrm{d}\sigma}
		\right)^2
		+
		\frac
			{r^4}
			{L^4}
		\left(
			1-\frac
				{r_h^4}
				{r^4}
		\right)
		+
		r^2\dot{\phi}^2~
		}
	& = &
	C_1,~~~~~~
\end{eqnarray}
Thus,
\begin{eqnarray}\label{3-21}
	-
	\frac
		{
			\frac
				{r^4}
				{L^4}
			\left(
				1-\frac
					{r_h^4}
					{r^4}
			\right)
			+
			r^2\dot{\phi}^2
		}
		{
			\sqrt{
				\left[
					1+
					\frac
						{L^4}
						{r^2}
					\left(
						1-\frac
							{r_h^4}
							{r^4}
					\right)^{-1}
					\dot{\phi}^2
				\right]
				\left(
					\frac
						{\mathrm{d}r}
						{\mathrm{d}\sigma}
				\right)^2
				+
				\frac
					{r^4}
					{L^4}
				\left(
					1-\frac
						{r_h^4}
						{r^4}
				\right)
				+
				r^2\dot{\phi}^2~
				}~
		}
	& = &
	C_1.
\end{eqnarray}
Considering the same boundary condition in zero temperature case Eq.\eqref{2-12}, and substituting it into Eq.\eqref{3-21}, we can find the conserved constant
\begin{eqnarray}\label{3-24}
	C_1
	& = &
	-\sqrt{
		\frac
			{r_c^4}
			{L^4}
		\left(
			1-\frac
				{r_h^4}
				{r_c^4}
		\right)
		+r_c^2\dot{\phi}^2~
	}.
\end{eqnarray}
Putting the Eq.\eqref{3-24} into Eq.\eqref{3-21}, we obtain the equation for finite temperature, that is,
\begin{eqnarray}\label{3-26}
	\frac
		{\mathrm{d}r}
		{\mathrm{d}\sigma}
	& = &
	\frac
		{1}
		{L^2}
	\sqrt{
		\frac
			{
				(r^2-r_c^2)
				(r^4-r_h^4)
				(r^2+r_c^2+L^4\dot{\phi}^2)~
			}
			{
				r_c^4-r_h^4+L^4\dot{\phi}^2}
			}.
\end{eqnarray}
\begin{figure}[ht]
	\centering
	\includegraphics[width=10cm]{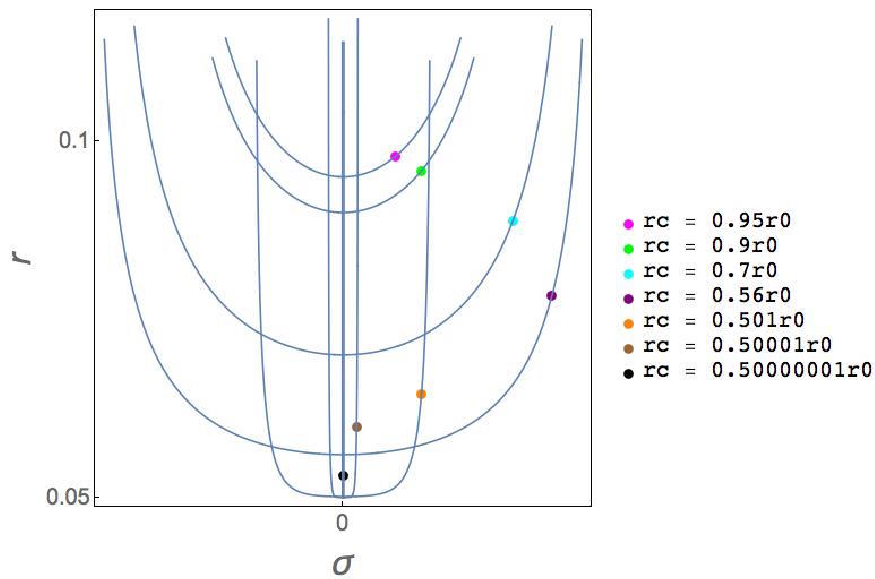}
	\caption{Choosing $r_h = 0.05 $ and $r_0 = 0.1$, the different positions of $r_c $ will lead to the different shapes of D3-brane at finite temperature, as shown in Fig.\ref{fig-6}. When the horizon $r_h$ gets closer to $r_c$, that is, the higher the temperature, the D3-brane configuration will become longer and thinner, the separate distance $x$ will become smaller. When $r_h$ closes to $r_c$ infinitely, we can see the D3-brane be drawn into a thin line with zero separate length.}
	\label{fig-6}
\end{figure}

In order to arrive at Eq.\eqref{3-26}, we must first consider the situation of $\dot{\phi}=0$. Eq.\eqref{3-2} says that the temperature is proportional to the radius of horizon $r_h$, and thus that choosing a fixed temperature is equivalent to choosing a fixed value of $r_h $. Because the position of probe D3-brane has been fixed at $r = r_0$, the rest parameters are $r_c$ and $r(\sigma)$. In the other words, when we choose a vertex $r_c $ of the D3-brane, which can draw a configuration figure of about $r (\sigma)$. Choosing $r_h = 0.05 $ and $r_0 = 0.1$, the different positions of $r_c $ will lead to different shapes of D3-brane at finite temperature, as shown in Fig.\ref{fig-6}. When the horizon $r_h$ gets closer to $r_c$, that is, the higher the temperature, the D3-brane configuration will become longer and thinner, the separate distance $x$ will become smaller. When $r_h$ closes to $r_c$ infinitely, we can see the D3-brane be drawn into a thin line with zero separate length.

When D3-brane is drawn into a thin line, which represents the spacing is always zero, it means that no matter where the probe point $r_0 $ is in any position, the distance of test particle is always zero. Back to the previous section for the total potential energy equation \eqref{3-30}, the distance of test particles is identified by $x = 0 $. It implies that the virtual particles can't get any energy from the external electric field. Particles and anti-particles are firmly locked together, never separate and never inspire real particles. This is a failure of the Schwinger effect because the particles will remain at annihilated state, which is an absolute vacuum state.

About the separate distance $x$, integrating Eq.\eqref{3-26}, we have
\begin{eqnarray}\label{3-27}
	\mathlarger{x}
	& = &
	 \mathlarger{2\int_{r_c}^{r_0}}
		\mathrm{d}r~
		L^2
		\sqrt{
		\frac
			{
				r_c^4-r_h^4+L^4\dot{\phi}^2
			}
			{
				(r^2-r_c^2)
				(r^4-r_h^4)
				(r^2+r_c^2+L^4\dot{\phi}^2)~
			}~
		}.
\end{eqnarray}

\begin{figure}[ht]
	\centering
	\includegraphics[width=8cm]{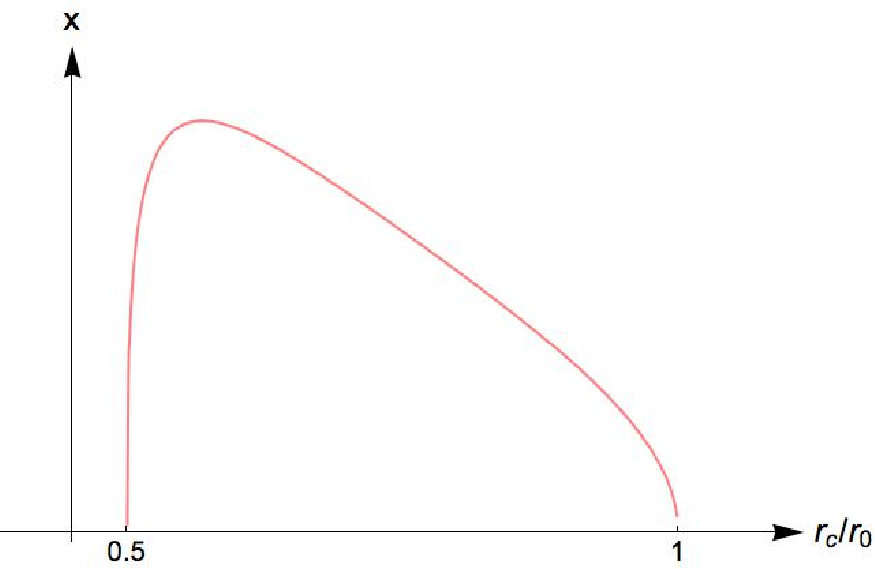}
	\caption{The relation between different $r_c$ and the separate length of the test particles on the rotating probe D3-brane when $r_h=0.05$ and $r_0=0.1$ at finite temperature. The separate length $x$ decays to zero rapidly when $r_c$ gradually gets closer to $r_h=0.05$.}
	\label{fig-7}
\end{figure}

The result is shown in Fig.\ref{fig-7}. We can find that the separation distance of the test particles first gets a similar linear growth from $r_c = r_0$. Because of the black hole's gravitation, $x $ decays to zero rapidly when $r_c $ crosses a certain point and gradually closes to $r_h$ \citep{rey1998wilson}.

\begin{figure}[!ht]
	\centering
	\subfigure[]
	{
		\includegraphics[width=0.43\textwidth]
			{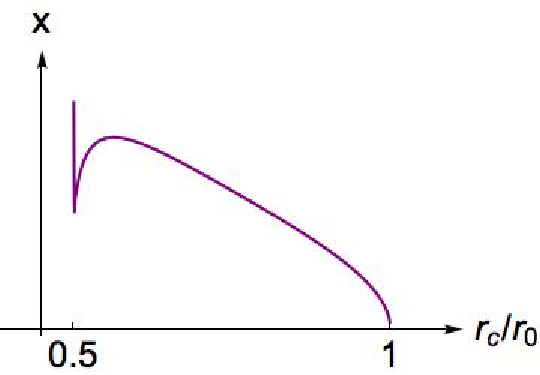}
	}
	~~~~
	\subfigure[]
	{
		\includegraphics[width=0.43\textwidth]
			{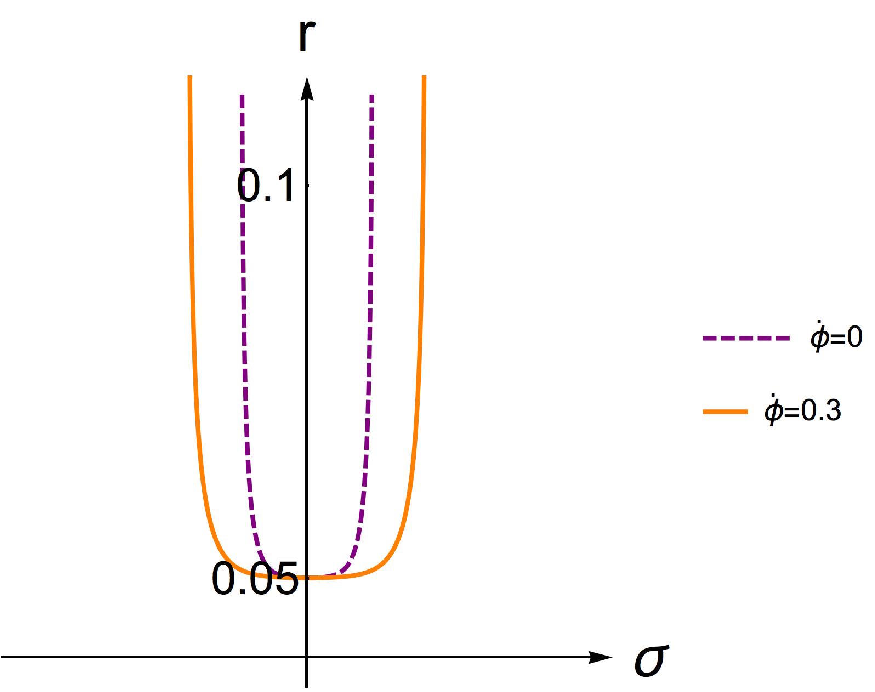}
	}
	\caption{ (a) There is a rapid increase near $r_c = r_h$ significantly. (b) Both dotted and solid lines, their vertices both are at $r_c =0.501r_0$. Because there is the angular velocity $\dot{\phi} = 0.3$, the solid line "open"  a wider width than $\dot{\phi} = 0$.}
\label{fig-8}
\end{figure}

Now we consider the situation where $\dot{\phi}$ is not zero. Suppose $\dot{\phi}=0.3$, we re-examine Eq.\eqref{3-26} and the test particle separation distance \eqref{3-27}, the results are shown in Fig.\ref{fig-8}. It can be seen in Fig.\ref{fig-8} (a), there is a rapid increase near $r_c = r_h$ significantly. In Fig.\ref{fig-8} (b), the vertices of both the dotted and solid lines are at $r_c =0.501r_h$. Because of the angular velocity $\dot{\phi} = 0.3$, the solid line "open"  a wider width than $\dot{\phi} = 0$.

As shown in Fig.\ref{fig-6}, if D3-brane is drawn into a thin line where the spacing is zero (absolute vacuum state), the rotation in $S^5$ is a way to break the absolute vacuum state. Fig.\ref{fig-8} (a) clearly shows that once an angular velocity $\dot{\phi}$ exists, the length of test particles is increased significantly near the horizon, preventing the emergence of an absolute vacuum. This rotation is just an objective physical phenomenon.

When $r_h$ gets further away from $r_c $ or $r_h $ goes further away from $r_0$ near the horizon, the separate distance of test particles will increase. Similarly, when we take an angular velocity $\dot{\phi}$ in $S^5$ spherical, the separate length spacing will also increasing. Through Eq.\eqref{3-2} we realize that the smaller $r_h $ means the lower the temperature, and this cooling behavior corresponds to a kind of spherical rotation. That is an interesting thing.

We would like to go over the world-sheet topology and the termination on the D3-brane again. Actually, the word-sheet topology is that of a disk with the boundary connected to the D3-brane.

In the usual studies, the test particles are assumed to be infinitely heavy. However, this assumption is not appropriate for considering the Schwinger effect in a holographic way because the pair creation is severely suppressed due to the divergent mass. Therefore, the setup should be modified so that the production rate makes sense. It is to put a probe D3-brane at an intermediate position $r=r_c$ rather than close to the boundary. The mass $m$ then becomes finite and depends on $r_c$, as $m=T_{F}r_c$, where $T_{F}=1/(2\pi\alpha^{\prime})$ denotes fundamental string tension.

\subsection{Total potential}
~~~~Now, at a finite temperature, we compute the total potential by turning on an electric field $E$ along the $x^1$ direction. For simplicity, we introduce a dimensionless parameter $\alpha$ \eqref{2-29}.
So the total potential $V_{tot}$ is written as
\begin{eqnarray}\label{3-30}
	V_{tot}
	& = &
	V_{\mathrm{CP+E}}-Ex
	\nonumber \\
	& = &
	2~T_F
	\Bigg[
	\mathlarger{\int_{r_c}^{r_0}}
		\mathrm{d}r~
		\frac
			{r^4-r_h^4+L^4r^2\dot{\phi}^2}
			{
			  \sqrt{
			  	(r^2-r_c^2)
				(r^4-r_h^4)
				(r^2+r_c^2+L^4\dot{\phi}^2)
			  ~}~
			}
	\nonumber \\
	\nonumber \\
	&-&\alpha
	\frac
		{r_0^2}
		{L^2}
	\sqrt{
		1
		-
		\frac
			{r_h^4}
			{r_0^4}
		+\frac
			{L^4}
			{r_0^2}
		\dot{\phi}^2
	}
	 \mathlarger{\int_{r_c}^{r_0}}
	 	\mathrm{d}r
		L^2
		\sqrt{
		\frac
			{
				r_c^4-r_h^4+L^4\dot{\phi}^2
			}
			{
				(r^2-r_c^2)
				(r^4-r_h^4)
				(r^2+r_c^2+L^4\dot{\phi}^2)
			}
		}
	\Bigg].~~~~~~~~
\end{eqnarray}
\begin{figure}[ht]
	\centering
	\includegraphics[width=6cm]{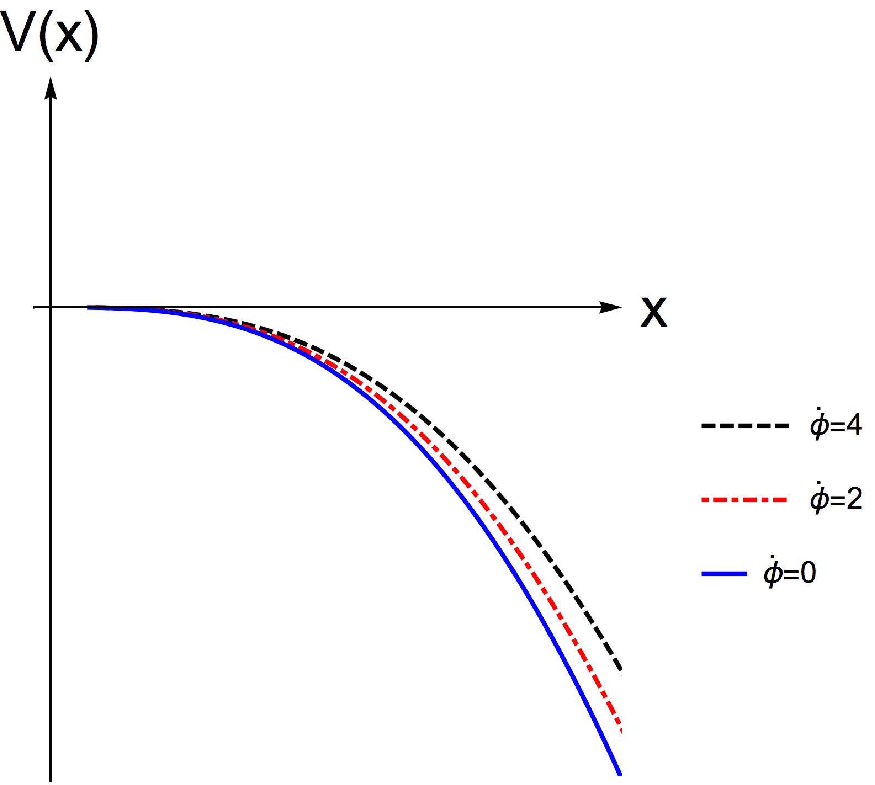}
	\caption{For $\alpha=1$, when the angular velocity $\dot{\phi}=0,2,4$, respectively, the relation between the total potential $V_{tot}$ and the separate distance $x$ is shown. It can be seen that the potential barrier just vanishes, so the result completely agrees with the DBI result.}
	\label{fig-9}
\end{figure}
\begin{figure}[!ht]
	\centering
	\subfigure[]{\includegraphics[width=6cm]{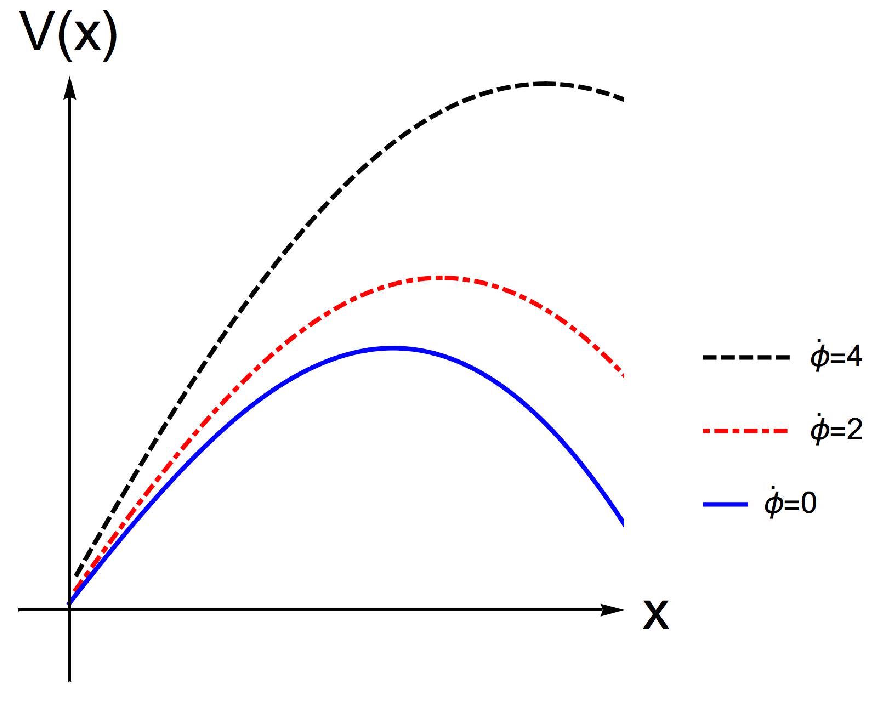}}~~~~ \subfigure[] {%
	\includegraphics[width=6cm]{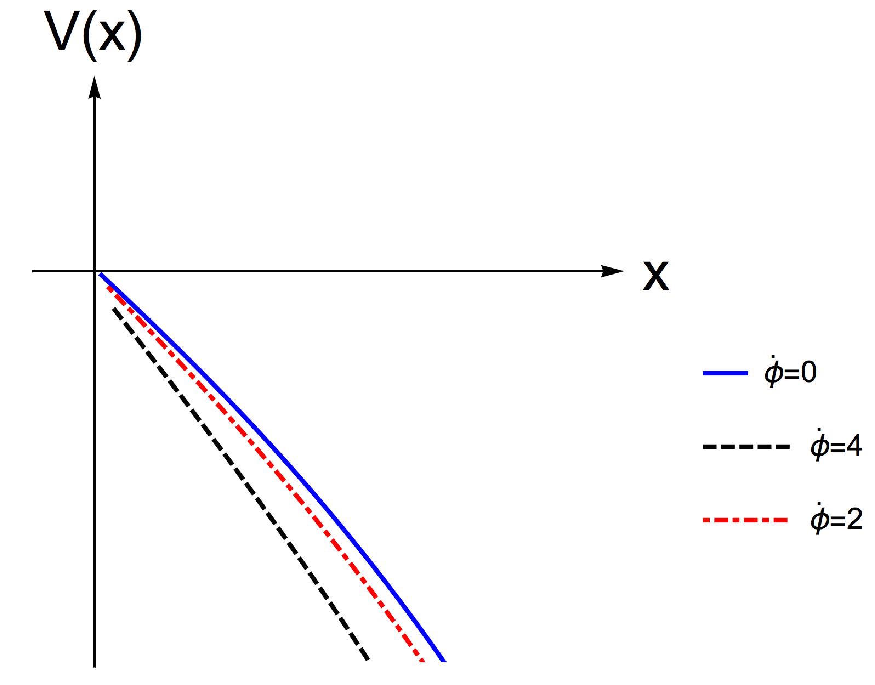}}
	\caption{(a) When $\alpha=0.8$ ($E<E_c$), the potential barrier exists, the real particle production rate is exponentially suppressed, and the vacuum is stable. (b) When $\alpha=1.2$ ($E>E_c$), the potential barrier has already vanishes. The vacuum becomes totally catastrophic unstable.}
	\label{fig-10}
\end{figure}
When $\alpha=1$, if the angular velocity $\dot{\phi}=0$, $\dot{\phi}=2$ and $\dot{\phi}=4$ respectively, the relation between the total potential $V_{tot}$ and the separate distance $x$ is shown in Fig.\ref{fig-9}. It can be seen that the potential barrier just vanishes, so the result agrees completely with the DBI result. For $\alpha=0.8$ ($E<E_c$) and $\alpha=1.2$ ($E>E_c$), the results are showed in Fig.\ref{fig-10}.

\section{Discussion, Summary and Conclusion}
~~~~We fix the probe D3-brane in the middle of $AdS_5\times S^5$ to have the angular velocity and then proceed to finish the potential investigation in holographic Schwinger effect. We discover that for zero temperature case, the faster the angular velocity, the farther the distance of the test particles at D3-brane, the potential barrier of total potential energy also grows higher and wider. For finite temperature case, besides the potential research under AdS black hole metric, we find that the system temperature is associated with a certain direction angular velocity in $S^5$ in D3-brane configuration. We further discover that Near the horizon, if the system temperature is reduced, it means that the gravity effect grows weak, the separate distance of test particles increases and a certain angular velocity in $S^5$ can achieve the same effect.

There have been past studies on the Schwinger effect in higher-dimensional theories, and it is important to compare and contrast the new results with these past studies.
One significant difference between our study and past studies is the use of holography to study the effect. Holography is a powerful tool in string theory that relates gravitational theories in higher dimensions to lower-dimensional field theories. By using holography,  we were able to study the Schwinger effect on the D3-brane using the DBI action, which is the low-energy effective action of the D3-brane in string theory. This allowed us to study the effect in a regime where traditional field theory methods are not applicable.
Another significant difference is the inclusion of rotation of the D3-brane in the analysis. We showed that the rotation of the D3-brane can significantly modify the potential barrier for the Schwinger effect, allowing it to occur even at finite temperature where it would otherwise be suppressed. This is a new result that has not been studied in previous works on the Schwinger effect in higher-dimensional theories.
Furthermore, we investigated the effect of the distance and angular velocity of the test particle pair on the potential barrier, which provides new insights into the behavior of the Schwinger effect in the presence of rotation. This is important for understanding the fundamental physics underlying the brane world picture and string theory.
In summary, the use of holography, the inclusion of rotation, and the investigation of the effect of distance and angular velocity on the potential barrier are significant new contributions of our work compared to past studies on the Schwinger effect in higher-dimensional theories. These new results provide further insights into the behavior of the Schwinger effect in the presence of rotation and can help us better understand the fundamental physics underlying string theory and the brane world picture.

The Dirac-Born-Infeld (DBI) action is a theoretical framework that describes the behavior of D-branes in string theory. D-branes are objects that arise in string theory and can be thought of as surfaces on which strings can end. The DBI action describes the dynamics of these surfaces in terms of their position, velocity, and other properties.
The fact that the results of the Schwinger effect on the D3-brane, both at zero and finite temperature, agree with the predictions of the DBI action is significant because it provides further evidence for the validity of the brane world picture and string theory in describing high-energy physics.
String theory and the brane world picture predict the existence of extra dimensions beyond the familiar four dimensions of space-time. They also predict the existence of new particles and interactions that are not part of the Standard Model of particle physics. The DBI action, which is a key component of these theories, provides a way to describe the dynamics of these extra dimensions and particles.
By studying the behavior of the D3-brane in the presence of an electric field, and by comparing the results with the predictions of the DBI action, we can gain insight into the fundamental physics underlying the brane world picture and string theory. This can help us to better understand the behavior of the universe at a fundamental level and to develop new theories and technologies based on these insights.

Generally speaking, the motion, especially rotation, will correspond to a certain way of doing work, and it will inevitably produce heat. However, in our model, which is constructed by the rotating probe D3-brane at a finite temperature, if there is an angular velocity in $S^5$, the model can carry the centrifugal force, which will make the distance between the test particles pair farther. This process is equivalent to the decrease of horizon $r_h$  (it means that the gravity effect grows weaker) and is equivalent to the cooling process of the system temperature.

This paper, for the first time, discover that at finite temperature, when $S^5$ without rotation, because D3-brane can be drawn into a zero-distance line of test particles pair, which leads to the occurrence of an absolute vacuum, this is a failure of the Schwinger effect because the particles will remain at annihilate state; but once there is the angular velocity in $S^5$, the increase of the test particle's distance will avoid the existence of an absolute vacuum near the horizon.

After setting the probe D3-brane to have the angular velocity, for both zero temperature and finite temperature states,  the critical field $E_c$ we obtained in the potential investigation completely agrees with the results of the DBI action. And in terms of the whole investigation in this paper, people can believe that this model will play important role in pair production research in the future.
\section*{Data Availability}
The data used to support the findings of this study are included within the article.
\section*{Acknowledgment}
The work is supported by National Natural Science Foundation of China (Grants No. 11875081).

\bibliographystyle{unsrtnat}
\bibliography{references}
\end{document}